\documentclass[%
 preprint,onecolumn,
 amsmath,amssymb,
 aps,
]{revtex4-1}

\usepackage{graphicx}
\usepackage{dcolumn}
\usepackage{bm}

\begin{document}


\title{Time variation of the equation of state for dark energy}

\author{Tetsuya Hara}
\email{hara@cc.kyoto-su.ac.jp}
\author{Ryohei Sakata}%
\altaffiliation{Altech Corporation, Queen's Tower C 18F, Yokohama, 220-6218, Japan}
\author{Yusuke Muromachi} 
\author{Yutaka  Itoh}
\email{yitoh@cc.kyoto-su.ac.jp}
\affiliation{Department of Physics, Kyoto Sangyo University, Kyoto 603-8555, Japan
}%
\begin{abstract}
The time variation of the equation of state ($w_Q$) for the dark energy is 
analyzed by the current values of parameters  $\Omega_Q $, $w_Q $ and their time derivatives. In the future, detailed features of the dark energy could be observed,
so we have considered the second derivative of $w_Q$  for two types of potential: One is an inverse power-law type ($V=M^{4+\alpha}/Q^{\alpha}$) and the other is an exponential one 
($V=M^4\exp{(\beta M/Q)}$).   
The first derivative $dw_Q/da$ and the second derivative $d^2 w_Q/da^2$ for both potentials are derived.  The first derivative is estimated 
by the observed two parameters $\Delta=w_Q+1$ and $\Omega_Q$, with assuming for $Q$.  
In the limit $\Delta \rightarrow 0$, the first derivative is null and, under the tracker approximation, the second derivative also becomes null. \\ 
{\indent} The evolution of forward and/or backward time variation could be analyzed from some fixed time point. If the potential is known, the evolution will be estimated from values $Q$ 
and $\dot {Q}$ at this point, because the equation for the scalar field is the second derivative equation.  
For the inverse power potential, 
if we do not adopt the tracker approximations, 
the observed first and second derivatives with $\Delta$ and $\Omega_Q$ must be utilized to determine the two parameters of the potential, $M$
 and $\alpha$.  For the exponential potential, 
  the second derivative is estimated by the observed parameters $\Delta$, $\Omega_Q$ and $dw_Q/da$,  because the parameter for this potential 
 is assumed essentially one, $\beta.$  If the parameter number is $ n$ for the potential form, it will be necessary for $n+2$ independent observations to determine 
the potential, $Q$ and $\dot{Q}$, for the evolution of the scalar field. 
\end{abstract}

\maketitle


\section{Introduction}

   Even though it is almost one and a half decadesl since the detection of the acceleration of the universe, the dark energy is not well understood \cite{1}.
We do not yet know whether it is the cosmological constant or not \cite{2,2-a, 3}.  Then a search is undertaken to observe 
the variation of the equation of state ($w=w_Q$) for the dark energy.  
Many works have been done on the study of dark energy in the form of a slowly rolling scalar field and
time variation of the equation of state for the dark energy. 
Usually, parameters are taken to denote the variation of $w$ as \  \cite{1,4,5,6,7,8,9,10}
\vspace{-0.6cm}

\begin{align}
 w(a)=w_0+w_a(1-a) , \label{act0}
\end{align}
where  $a, w_0, $ and $ w_a$ are the scale factor ($a=1$ at present), the current value of $ w(a), $ and the first derivative of $ w(a)$ by $w_a=-dw/da$, respectively.   

  Even now, not very much is known about the values of $w_0$ and $w_a$; we extend the parameter space as 
\begin{align}
w(a)=w_0+w_a(1-a)+\frac{1}{2}w_{a2}(1-a)^2 , \label{act1}
\end{align}
including the second derivative of $w(a)$ as $w_{a2}=d^2 w/da^2$. 
Although it may be hard to observe the parameter $w_{a2}$, it must be a good clue to understanding the features of dark energy in the future. 
One of the new ingredients of this work in comparison with past works is the inclusion of the second derivative for the parameter space, 
where $ w_Q$ and $dw_Q/da$ were mainly considered up to now \cite{11}. 

We follow the single scalar field formalism of Ref. \cite{12}  and take the potential of two types as $ V=M^{4+\alpha}/Q^{\alpha}$ and 
$  V=M^4\exp(\beta M/Q)$.  
Although there are many scalar potentials motivated by particle physics \cite{13,14}, the inverse power-law type is one of the most commonly investigated; 
it was chosen by Peebles and Ratra \cite{2,2-a} for the time variable 
``cosmological constant".  It is also examined in Ref. \cite{12} as one of the leading candidates for the ``tracker solution", 
which is the attractor-like solution to the ``coincidence problem", converging a very wide range of initial conditions to a common evolutionary track of $w_Q(t)$.

Steinhardt et al. \cite{12} also studied the exponential of the inverse power-law model $V(Q)=M^4\exp(1/Q)$ as a more generic potential with a mixture of inverse power-laws, 
through changing the effective power index by varying $Q$ (initially $ Q$ is small with a high power index; later it is large with a low power index).  
 Wang et al. \cite{15} have analyzed the tracker field quintessence models with potentials $V(Q) \propto Q^{-\alpha}, \ \exp(M_p/Q),  \ \exp(M_p/Q)-1, \ \exp(\beta M_p/Q), 
 $ and $\exp(\gamma M_p/Q)-1,$
with observational data from the SNIaCCMB, and BAO.  They have found that models with potentials $\exp(M_p/Q),  \exp(M_p/Q)-1,$ and  $
 \exp(\gamma M_p/Q)-1,$  are not supported by the data.
  So we examine the inverse power-law and the exponential one for the single scalar field model of dark energy.
  
 The goal of this paper is to explore the dark energy under the quintessence model in a single scalar field by assuming the potential. 
 We have tried to increase the parameter space to explore the features of dark energy, by adding the second derivative.  
 We examine our trial under typical potentials.  
To determine the potential form we must observe the evolution of the universe. 
If the parameter number is $ n$ for the potential form, it will be necessary to make $n+2$ independent observations to determine 
the potential, $Q$ and $\dot{Q}$, at some time for the time variation of the scalar field.  
\\

In Sect. 2, the equation of state for the scalar field and parameters to describe the potentials are presented. 
The first derivatives of $w_Q$ for two types of potentials are calculated in Sect. 3.  In Sect. 4, the second derivatives are presented, 
and the detailed calculations are shown in Appendix A. 
The results and discussion are presented in Sect. 5.

\section{Equation of state}

 \subsection{Scalar field}
 
 For the dark energy, we consider the scalar field $Q(\bm{x},t)$, where the action for this field in the gravitational field is described by \  \cite{12} 
 \begin{align}
S=\int d^4 x \sqrt{-g} [ -\frac{1}{16 \pi G}R+ \frac{1}{2}g^{\mu \nu} \partial _ \mu Q \partial _ \nu Q -V( Q ) ] +S_M ,  \label{act2}
\end{align}
 where $S_M$ is the action of matter field and $G$ is the gravitational constant, occasionally putting $G=1$.
 Neglecting the coordinate dependence, the equation for $Q(t)$ becomes as
 \begin{align}
\ddot{Q} + 3H \dot{Q}+V'=0  , \label{Qfield}
\end{align}
where $H$ is the  Hubble parameter and $V'$ is the derivative of $V$ by $Q$.  As $\kappa=8\pi/3$, $H$ satisfies the following equation
\begin{align}
H^2=(\frac{\dot{a}}{a})^2=\kappa (\rho _B + \rho _ Q)=\kappa \rho_c  , \label{Friedmann}
\end{align}
where $\rho _B$, $\rho _ Q $ and $\rho_c$ are the energy density of the background, and scalar fields and the critical density of the universe.
The energy density and pressure for the scalar field are written as
\begin{align}
\rho _ Q =\frac{1}{2} \dot{Q}^2+V \label{density} ,   \end{align}
and
\begin{align}
p_ Q =\frac{1}{2} \dot{Q}^2-V  , \label{pressure}
\end{align}
respectively.  Then the parameter $w_Q$ for the equation of state is described by 
\begin{align}
w_ Q \equiv \frac{p_ Q}{\rho _ Q}=\frac{ \frac{1}{2} \dot{Q}^2-V}{ \frac{1}{2} \dot{Q}^2+V} . \label{state}
\end{align}

 \subsection{Time variation of $w_Q$}
 It is assumed that the current value of $w_Q$ is slightly different from a negative unity by $\Delta ( > 0)$ as
 \begin{align}
w_Q=-1+\Delta     .\label{w}
\end{align}
By using Eq.(\ref{state}), $\dot{Q}^2$ is written as 
\begin{align}
\dot{Q}^2=\frac{2\Delta V}{2-\Delta} ,  \label{(2)}
\end{align}
which is also given by using the density parameter $\Omega_Q=\rho_Q/\rho_c$ as
 \begin{align}
\dot{Q}^2=2(\rho_c\Omega_Q-V) . \label{(3)}
\end{align}
Combining Eqs.(\ref{(2)}) and (\ref{(3)}), $V$ is given by
\begin{align}
V=\rho_c\Omega_Q(1-\frac{\Delta}{2}) . \label{potential}
\end{align}
From Eqs.(\ref{(3)}) and (\ref{potential}), $\dot{Q}$ is given as
\begin{align}
\dot{Q}=\sqrt{\Delta(\rho_c\Omega_Q)}.
\end{align}

If we determine the potential, the parameters to describe the evolution of the scalar field are the values of $Q$ and $\dot{Q}$ at some fixed time,
 because Eq.(\ref{Qfield}) is the second derivative equation.  Then the evolution or backward variation could be estimated from this fixed point.  
  In the following we take this fixed time as it is at present and estimate backward the accelerating behavior 
 in the near past.

 As $\rho_c$ is given by observation through the Hubble parameter $H$, 
 $\dot{Q}$ is determined by  $\Omega_Q$ and $\Delta$, which also determine the value of $V$.  If we adopt the form and parameter of the potential, 
 the value of $V$ could be used to estimate the value of $Q$. 
 In reality, the evolution of $H$ in Eq. (\ref{Qfield}) depends on the background densities, which include radiation density.  
 The effect of radiation density can be ignored in the near past ($z \leq 10^3$) and so is not considered in this work.

In the following, we investigate the power inverse potential $V=M^4(M/Q)^{\alpha} ; (\alpha >0) $ and
 the exponential potential $V=M^4\exp(\beta M/Q)  ; (\beta >0)$, respectively.
\vspace{0.1cm}

\subsubsection {  $V=M^4(M/Q)^{\alpha} $}
The parameters of the potential are $M$ and $\alpha$. When we take $M=M_{\ast}$, $ V$ becomes
\begin{align}
V=M^4_{\ast}(\frac{M_{\ast}}{Q})^{\alpha} . \label{pote1}
\end{align}
From Eq. (\ref{potential}) as $M_{\ast}^4(M_{\ast}/Q)^{\alpha} =\rho_c\Omega_Q(1-\frac{\Delta}{2})$, $Q$ is given as
\begin{align}
 Q=\left(\frac{M_{\ast}^{4+\alpha}}{\rho_c\Omega_Q(1-\frac{\Delta}{2})}\right)^{1/\alpha} .\label{(4)}
\end{align}
If we take $Q=Q_0M_{\rm pl}$ at present, $M_{\rm pl}$ being the Planck mass, $M_{\ast}$ becomes
\begin{align} 
M_{\ast}= M_{\rm pl}\left(Q_0^{\alpha}\frac{\rho_c}{M_{\rm pl}^4}\Omega_Q (1-\frac{\Delta}{2})\right)^{1/(4+\alpha)} .
\end{align}
Then $Q_0, \Omega_Q, \Delta$, and $\alpha$ determine the parameter $M_{\ast}$, which means that the parameters determining the accelerating behavior are 
$Q_0, \Omega_Q, \Delta$, and $\alpha$.

The difference in the observed value $\rho_c$ and the value $M^4_{\rm pl}$ is described by the observed value $N$ as
\begin{align}
\rho_c = M^4_{\rm pl} \times 10^{-N} ;  \label{4.01}
\end{align}
as $N\simeq 122$ \cite{3},  $M_{\ast}$ becomes 
\begin{align} 
M_{\ast}= M_{\rm pl} \times \left(Q_0^{\alpha}\Omega_Q (1-\frac{\Delta}{2})\right)^{1/(4+\alpha)}\times 10^{-N/(4+\alpha)} .\label{3.91}
\end{align}
The problem is how to estimate $Q_0$ and $\alpha$.

\subsubsection{$V=M^4\exp(\beta M/Q)$ }
If we put $M_{\rm pl}=\beta M$, $V$ is written as
\begin{align}
V=\left(\frac{M_{\rm pl}}{\beta}\right)^4\exp \left(\frac{M_{\rm pl}}{Q}\right). \label{exp}
\end{align}
In essence, as $\beta$ is combined with $M$, the parameter of this potential is $\beta$.
From Eq.(\ref{potential}),  $\exp(M_{\rm pl}/Q)$ is given by
\begin{align}
\exp\left(\frac{M_{\rm pl}}{Q}\right)=\left(\frac{\beta}{M_{\rm pl}}\right)^4\rho_c\Omega_Q(1-\frac{\Delta}{2})  ,
\end{align}
Then $Q$ is estimated as
\begin{align}
Q=\frac{M_{\rm pl}}{\ln(\beta^4\rho_c\Omega_Q(1-\frac{\Delta}{2})/M^4_{\rm pl})}. \label{expQ}
\end{align}
\vspace{0.2cm}

If we take $Q=Q_0 M_{\rm pl}$ at present, $Q_0$ determines the parameter $\beta$ as,
\begin{align}
\beta=\left(\frac{M^4_{\rm pl}}{\rho_c \Omega_Q(1-\Delta/2)}\right)^{1/4}\exp \left( Q^{-1}_0 /4 \right) .  \label{Vbeta}
\end{align}
In this potential, the parameters determining the accelerating behavior are $Q_0, \Omega_Q $, and $ \Delta$.  
The problem how to estimate $Q_0$ is also left.

\section{First derivative of $w_Q$}
To investigate the variation of $w_Q$, we calculate $dw_Q/da$, using Eqs. (\ref{Qfield}), (\ref{density}) and (\ref{pressure}),
\begin{align}
\frac{dw_Q}{da} & = \frac{1}{\dot{a}}\frac{d}{dt}\left(\frac{p_Q}{\rho_Q}\right)=\frac{1}{\dot{a}}\frac{\dot{p}_Q\rho_Q-p_Q\dot{\rho}_Q}{\rho^2_{Q}} \nonumber \\
& = \frac{1}{aH\rho^2_Q}(\dot{Q}(\ddot{Q}-V')\rho_Q-p_Q\dot{Q}(\ddot{Q}+V')) \nonumber \\
& = \frac{\dot{Q}}{aH\rho^2_Q}((-3H\dot{Q}-2V')\rho_Q-p_Q(-3H\dot{Q})) \nonumber  \\
& = \frac{\dot{Q}}{aH\rho^2_Q}(3H\dot{Q}(p_Q-\rho_Q)-2V'\rho_Q) \nonumber \\
& = \frac{2V\dot{Q}}{aH\rho^2_Q}\left(-3H\dot{Q}-\frac{V'}{V}\rho_Q\right) . \label{dw/da}
\end{align}
In the limit $\Delta \rightarrow 0$ where $\dot{Q}=\sqrt{\Delta \rho_c \Omega_Q} \rightarrow 0$, $dw_Q/da$ becomes null.
To investigate further, we must consider the potential form.
\vspace{0.2cm}

\subsection{  $ V=M^{4+\alpha}/Q^{\alpha}$ }

As $V'/V=-\alpha/Q $, Eq.(\ref{dw/da}) becomes  
\begin{align}
\frac{dw_Q}{da} & = \frac{2V\dot{Q}}{aH\rho^2_Q}\left(\frac{\alpha\rho_Q}{Q}-3H\dot{Q}\right) \nonumber \\
& = \frac{2V\dot{Q}}{aH Q\rho_Q}\left(\alpha-\frac{3H\dot{Q}Q}{\rho_Q}\right) . \label{4.14}
\end{align}
To investigate the signature of $dw_Q/da$, we must estimate the following term;
\begin{align}
\frac{3H\dot{Q}Q}{\rho_Q} & = \frac{3H}{\rho_c\Omega_Q}\sqrt{\Delta(\rho_c\Omega_Q)}Q_0M_{\rm pl}=\sqrt{\frac{24\pi\Delta}{\Omega_Q}}\times Q_0.  \label{(6)}
\end{align}
To estimate $Q_0$, we consider the tracker approximation that $w_Q$ is almost constant as
 \begin{align}
 -\sqrt{\frac{\Omega_Q}{24\pi(1+w_Q)}}\frac{V'}{V}=1,                                                                         \label{(6c)}
 \end{align}
which is given by Eq. (9) of  Steinhardt et al. \cite{12}
If we adopt this approximation, $Q/M_{\rm pl}$ becomes as 
 \begin{align}
 Q_{\alpha}=\sqrt{\frac{\Omega_Q}{24\pi\Delta}}\times \alpha .                                                    \label{QA}
 \end{align}
We approximate the present value $Q_0$ as $Q_0=(1+\varepsilon)\times Q_{\alpha}$.  
Then from Eq. (\ref{(6)}), $\sqrt{\frac{24\pi\Delta}{\Omega_Q}} \times Q_0$ becomes $(1+\varepsilon)\alpha$.
If $\varepsilon > 0$, $dw_Q/da < 0$, vice versa.

 From Eq. (\ref{4.14}), $Q$ is derived as
 \begin{align}
 Q=\frac{2  \alpha  \rho_Q V \dot{Q}}{H(a\rho_Q^2\frac{dw_Q}{da}+6V\dot{Q}^2)},                                         \label{(6d)}
 \end{align}
 then $Q_0$ is given by 
 \begin{align}
 Q_0=\alpha \sqrt{\frac{3\Delta \Omega_Q}{2\pi}}\frac{(1-\Delta/2)}{(\frac{dw_Q}{da}+6\Delta(1-\Delta/2))}.              \label{(6e)}
 \end{align}
 If $dw_Q/da$ is observed, $Q_0/\alpha$ will be  determined by the observed values $\Omega_Q, \Delta, $ and $dw_Q/da$.
 
 For a positive value of $Q_0$, $\frac{dw_Q}{da}$ must be greater than  $-6\Delta(1-\Delta/2)$.
 Assuming that $dw_Q/da \lesssim 1$ and $\Delta \lesssim 0.1$, it is estimated that $ 0 \lesssim Q_0 \lesssim \infty$.

\subsection{$V=\left(\frac{M_{\rm pl}}{\beta}\right)^4\exp(\frac{M_{\rm pl}}{Q})$}
As $V'/V=-M_{\rm pl}/Q^2$, Eq.(\ref{dw/da}) becomes 
\begin{align}
\frac{dw_Q}{da} & = \frac{2V}{aH\rho_Q}\left[-\frac{3H\dot{Q}^2}{\rho_Q}-\frac{V'}{V}\dot{Q}\right] 
= \frac{2V}{a\rho_Q}\left[-\frac{3\dot{Q}^2}{\rho_Q}-\frac{V'\dot{Q}}{VH}\right]   \nonumber \\
& = \frac{2V}{a\rho_Q}\left[-\frac{3\rho_c\Delta \Omega_Q}{\rho_Q}+\frac{M_{\rm pl}}{Q^2}\frac{\sqrt{\Delta\Omega_QM^4_{\rm pl}
\times10^{-N_c}}}{\sqrt{8\pi/3}M_{\rm pl}\times10^{-N_c/2}}\right] \nonumber \\
& = \frac{2V}{a\rho_Q}\left[-3\Delta+\frac{\sqrt{\Delta\Omega_Q}}{(Q/M_{\rm pl})^2\sqrt{8\pi/3}}\right] \nonumber\\
& = \frac{2V}{a \rho_Q (Q/M_{\rm pl})^2}\sqrt{\frac{3\Delta\Omega_Q}{8\pi}}\left[1-\sqrt{\Delta/\Omega_Q}(Q/M_{\rm pl})^2\sqrt{24\pi}\right] .    \label{dw/daexp} 
\end{align}
If we adopt the tracker approximation in Eq. (\ref{(6c)}), $Q$ becomes
\begin{align}
Q_{\beta}=\left(\frac{\Omega_Q}{24\pi\Delta}\right)^{1/4} .    \label{QB} 
\end{align}
We approximate the current value $Q_0$ as $Q_0=(1+\varepsilon)\times Q_{\beta}$.  
Then from Eq. (\ref{dw/daexp}), the above value within [ \ \ \ ] becomes  $-\varepsilon$.
If $\varepsilon > 0$, $dw_Q/da < 0$,  and vice versa.

From Eq. (\ref{dw/daexp}), $Q$ is determined as 
 \begin{align}
 Q=\left(\frac{2V\rho_Q\dot{Q}}{H(a\rho^2_Q\frac{dw_Q}{da}+6V\dot{Q}^2)}\right)^{1/2}M_{\rm pl}.                                 \label{(6f)}
\end{align}
Then $Q_0$ is estimated by the observable parameters $\Omega_Q, \Delta,$ and $dw_q/da$ as
 \begin{align}
 Q_0=\left(\sqrt{\frac{3\Delta \Omega_Q}{8\pi}} \frac{2(1-\Delta/2)}{(a\frac{dw_Q}{da}+6\Delta(1-\Delta/2))}\right)^{1/2}. \label{(6g)}
\end{align}
$Q_0$ does not depend on the potential parameter $\beta$, which is determined by Eq. (\ref{Vbeta}).

For the real value of $Q_0$, $\frac{dw_Q}{da}$ must be greater than  $-6\Delta(1-\Delta/2)$.
 Assuming that $dw_Q/da \lesssim 1$ and $\Delta \lesssim 0.1$, it is estimated that $ 0 \lesssim Q_0 \lesssim \infty$.

\section{The second derivative of $w_Q$}
From Eq. (\ref{dw/da}), the second derivative of $w_Q$  is given by
\begin{align}
\frac{d^2w_Q}{da^2} & =\frac{1}{\dot{a}^3\rho_Q^4}[(\ddot{p}_Q\rho_Q-p_Q\ddot{\rho}_Q)\dot{a}\rho^2_Q  \nonumber \\
& -(\dot{p_Q}\rho_Q-p_Q\dot{\rho}_Q)(\ddot{a}\rho_Q^2+2\dot{a}\rho_Q\dot{\rho}_Q)]   \label{d^2w/da^2} . 
\end{align}
The time derivatives of $p_Q$ and $\rho_Q$ are written as
\begin{align}
\dot{p}_Q =-3H\dot{Q}^2-2V'\dot{Q}, \nonumber \\
\ddot{p}_Q=(-3\frac{\ddot{a}}{a}+21H^2-2V'')\dot{Q}^2+12H\dot{Q}V'+2(V')^2,  \label{5.2} \\
\dot{\rho}_Q =-3H\dot{Q}^2 , \ \ 
\ddot{\rho}_Q=(-3\frac{\ddot{a}}{a}+21H^2)\dot{Q}^2+6H\dot{Q}V' ,  \label{5.3} 
\end{align}
where $\ddot{a}/a=-4\pi G(\rho_c + p_Q)$.  
Using these equations, the detailed calculation of Eq.(\ref{d^2w/da^2}) is described in Appendix A.

From Eq.(\ref{4.56}), $d^2w/da^2 $ becomes 
\begin{align}
 & \frac{d^2w_Q}{da^2}= \frac{3}{4\pi}\frac{\Omega_Q}{a^2}\left(1-\frac{\Delta}{2}\right) \nonumber \\
& \times \biggl[-\Delta M^2_{\rm pl}\frac{V''}{V} +\sqrt{\frac{6\pi \Delta}{\Omega_Q}}\left(\left(1-\Delta\right)(6+\Omega_Q)-\frac{1}{3}\right)  \label{4.56a} \\
& \quad \times M_{\rm pl}\left(\frac{V'}{V}\right) +(1-\frac{\Delta}{2})M^2_{\rm pl}\left(\frac{V'}{V}\right)^2+\frac{8\pi\Delta}{\Omega_Q}(7-6\Delta) \biggr] .  \nonumber 
\end{align}
From this equation, the value of $d^2w_Q/da^2$ can be estimated.  
In the limit $\Delta \rightarrow 0$, the signature of $d^2w_Q/da^2$ is positive under the condition $V'/V \neq 0$.  
In principle, when the first derivative is observed, $Q_0$ is estimated through Eqs. (\ref{(6e)}) or (\ref{(6g)}).  
Then the second derivative is estimated by the observed parameters $\Delta$, $\Omega_Q$ and $Q_0$.

From this equation, we estimate $d^2w_Q/da^2$ for each potential in the following.

\subsection{$V=M^{4+\alpha}/Q^{\alpha}$}
As the following relations are derived
\begin{align}
\frac{V''}{V}=\frac{\alpha(\alpha+1)}{Q^2}, \ \frac{V'}{V}=-\frac{\alpha}{Q}, \ \left(\frac{V'}{V}\right)^2=\frac{\alpha^2}{Q^2}, \nonumber
\end{align}
we put them in Eq.(\ref{4.56a}) and  get
\begin{align}
\frac{d^2w_Q}{da^2} & = \frac{3}{4\pi}\frac{\Omega_Q}{a^2}\left(1-\frac{\Delta}{2}\right)
 \times \biggl[-\Delta \frac{\alpha(\alpha+1)}{(Q/M_{\rm pl})^2}-\sqrt{\frac{6\pi \Delta}{\Omega_Q}}\left(\left(1-\Delta\right)(6+\Omega_Q)-\frac{1}{3}\right)\left(\frac{\alpha}{Q/M_{\rm pl}}\right)  \label{4.57} \\
& \quad  +(1-\frac{\Delta}{2})\left(\frac{\alpha}{Q/M_{\rm pl}}\right)^2+\frac{8\pi\Delta}{\Omega_Q}(7-6\Delta) \biggr] .  \nonumber 
\end{align}
If $dw_Q/da$ is observed, $Q/\alpha$ will be determined by Eq. (\ref{(6e)}).  If $d^2w_Q/da^2$ is observed, one can estimate the value of $\alpha$ from the above equation.

When we take the tracker approximation as $Q_0\simeq (1+\varepsilon)\times Q_{\alpha} \simeq  (1+\varepsilon)\times \sqrt{\Omega_Q/(24\pi \Delta)}\times \alpha $ of  Eq. (\ref{QA}), 
the part within [ \ \ \ ] of the above equation becomes
\begin{align}
\frac{24\pi\Delta}{\Omega_Q}\left\{-\frac{\Delta(1+1/\alpha)}{(1+\varepsilon)^2}-\frac{1}{2(1+\varepsilon)}((1-\Delta)(6+\Omega_Q)-\frac{1}{3}\Omega_Q) 
  +\frac{1}{(1+\varepsilon)^2}(1-\frac{\Delta}{2})+\frac{1}{3}(7-6\Delta) \right\}   \label{4.57a}
\end{align}
In the limit $\Delta \rightarrow 0$ where $Q_{\alpha} \rightarrow \infty$, $dw^2_Q/da^2$ becomes null in this approximation.

\subsection{$V=(M_{\rm pl}/\beta)^4\exp(M_{\rm pl}/Q)$}

As the following relations are derived 

\begin{align}
\frac{V''}{V}=\frac{2M_{\rm pl}}{Q^3}+\frac{M^2_{\rm pl}}{Q^4}, \  \frac{V'}{V}=-\frac{M_{\rm pl}}{Q^2}, \ \left(\frac{V'}{V}\right)^2=\frac{M^2_{\rm pl}}{Q^4} . \nonumber
\end{align}
we put them in Eq.(\ref{4.56a}) and get
\begin{align}
\frac{d^2w_Q}{da^2} & = \frac{3}{4\pi}\frac{\Omega_Q}{a^2}\left(1-\frac{\Delta}{2}\right)
\times\biggl[-\Delta \left(2\left(\frac{M_{\rm pl}}{Q}\right)^3+\left(\frac{M_{\rm pl}}{Q}\right)^4\right) \nonumber \\
& \quad -\sqrt{\frac{6\pi \Delta}{\Omega_Q}}\left(\left(1-\Delta\right)(6+\Omega_Q)-\frac{1}{3}\Omega_Q\right)\left(\frac{M_{\rm pl}}{Q}\right)^2  \label{4.98b} \\
& \qquad  +(1-\frac{\Delta}{2})\left(\frac{M_{\rm pl}}{Q}\right)^4+\frac{8\pi\Delta}{\Omega_Q}(7-6\Delta) \biggr] .  \nonumber 
\end{align}

When we take the tracker approximation as $Q_0\simeq (1+\varepsilon)\times Q_{\beta}=(1+\varepsilon)\times (\Omega_Q/(24\pi \Delta))^{1/4}$ in Eq. (\ref{QB}), 
the part within [ \ \ \ ] of the above equation becomes
\begin{align}
& \biggl[\frac{24\pi\Delta}{\Omega_Q}\biggl\{-\Delta \left(\frac{2}{(1+\varepsilon)^3}(\frac{24\pi\Delta}{\Omega_Q})^{-1/4} 
 +\frac{1}{(1+\varepsilon)^4}\right)-\frac{1}{2(1+\varepsilon)^2}\left((1-\Delta)(6+\Omega_Q)-\frac{1}{3}\Omega_Q\right)    \nonumber \\    
& \quad +\frac{1}{(1+\varepsilon)^4}(1-\frac{\Delta}{2})+\frac{1}{3}(7-6\Delta )\biggr\} \biggr] .  \label{4.98c}
\end{align}
In the limit $\Delta \rightarrow 0$ where $Q_{\beta} \rightarrow \infty$, $dw^2_Q/da^2$ becomes null.

\section{Conclusions and discussion}
It is important to know the variation of the equation of state $w$  of the  background field for the investigation of the expansion of the universe.  
It is known that $\ddot{a}$ is described by the following equation
\begin{align}
\ddot{a}=-\frac{4\pi G}{3}(1+3w)a\rho .   \nonumber
\end{align}
At present, backward observation of large scale structure of the universe has been undertaken to estimate $w_Q$ at the age $(1+z)$ \cite{6}.
For the moment, the values of $w(a=a_0)$ and $dw/da$ have been pursued:
\begin{align}
w(a)=w(a=a_0)+\frac{dw}{da}da  .
\end{align}
Observation of the second derivative of $w$ could be expected in the future
\begin{align}
w(a)=w(a_0=1)+\frac{dw}{da}da+\frac{1}{2}\frac{d^2w}{da^2}(da)^2+\cdots  ,
\end{align}
so we estimate the second derivative of $w$ with $a$ for two typical potentials in this work.

The first derivative $dw_Q/da$ and the second derivative $d^2 w_Q/da^2$ for the power inverse and exponential potentials are calculated.  The first derivative is estimated 
by the two observed parameters $\Delta=w_Q+1$ and $\Omega_Q$, assuming parameters $Q_0$.  
In the limit $\Delta \rightarrow 0$, the first derivative is null and, under the tracker approximation, the second derivative also becomes null.  
For the inverse power potential $V=M_{\ast}^{4+\alpha}/Q^{\alpha}$, the observed first and second derivatives are used to determine the potential parameters $M_{\ast}$ and $\alpha$.  
For the exponential potential $V=M^4\exp{(\beta M/Q)}$, the second derivative is calculated by the observed parameters $\Delta$, $\Omega_Q$ and $dw_Q/da$, 
where the potential parameter is essentially one, $\beta$.

The evolution of forward and/or backward time variation could be analyzed from some fixed time point.  If the form of the potential is known, 
the evolution will be calculated with values $Q$ and $\dot {Q}$ at a fixed time,
 because the equation for the scalar field is the second derivative equation.  $\dot {Q}$ could be derived from $\Delta$ and $\Omega_Q$. 
If $dw_Q/da$ is observed, $Q/\alpha$ could be estimated for the inverse power-law potential and $Q$ could be estimated for the exponential potential. 
To estimate the parameter $\alpha$, one must observe the second derivative $d^2w_Q/da^2$.

After the parameters of the potential are determined, the time variation of the dark energy, such as $d^3w_Q/da^3, d^4w_Q/da^4 $ and so on, can be calculated.
If they are predicted values, it will be understood that the dark energy can be described by quintessence with a single scalar field.

We have increased the parameter space for dark energy such as $w$ and $dw/da$ by adding the second derivative $d^2w_Q/da^2$.  Then we have tried to derive the second derivatives 
for well investigated  potentials, suggested by the observations  \cite{15}.  Intimate relations between the second derivative and the observables have been derived.    
If we do not adopt some assumptions, such as tracker approximation and/or matter-dominant stage, 
it will be necessary to observe $ \Omega_Q, w_Q, dw_Q/da,$ and $ d^2w_Q/da^2 $ for the inverse power potential $V=M^4 (M/Q)^\alpha$, 
where the form parameters are two $M$ and $\alpha$.  When other forms of potential with parameter $n$ are taken, 
it is necessary to make independent $n+2 $ observations for the determination of $ Q, \dot{Q}$, 
and form parameter $n$ of the potential.  Such observations would be values of $\Omega_Q, w, dw/da, \cdots $ and $ d^nw/da^n$.

If we assume a matter-dominant approximation, we could estimate $\alpha$ and $M_{\ast}$ from the attractor solution \cite{16}, which is outlined in Appendix B. 
If $\Delta < 0$, we must consider utterly different models such as phantoms, quintom, or k-essence \cite{17, 18, 19}.

\vspace{1cm}
\hspace{-0.5cm} {\large{\bf Appendix}}

\appendix

\section{ Derivation of Eq. (\ref{d^2w/da^2}) }

By using Eqs. (\ref{5.2}) and (\ref{5.3}), the terms within Eq. (\ref{d^2w/da^2}) become 
\begin{align}
\hspace{-3cm}\ddot{p}_Q\rho_Q-p_Q\ddot{\rho}_Q & = [-2V''\dot{Q}^2+6H\dot{Q}V'+2(V')^2]\frac{1}{2}\dot{Q}^2 \nonumber \\
& \quad +[(-6\frac{\ddot{a}}{a}+42H^2-2V'')\dot{Q}^2+18H\dot{Q}V'+2(V')^2]V \label{4.31}\\
 & \qquad \dot{p}_Q\rho_Q-p_Q\dot{\rho}_Q  = \dot{Q}[3H\dot{Q}(p_Q-\rho_Q)-2V'\rho_Q]   . \label{4.33}
\end{align}
Next, we try to calculate  the second term within [ \ \ \ ] of Eq. (\ref{d^2w/da^2}).  It is given as \cite{16}
\begin{align}
\frac{\ddot{a}}{a}=4\pi G\left(-\frac{\rho_c}{3}-p_Q\right)  , \label{a}
\end{align}
where we neglect the radiation pressure.
By using the relation of $ p_Q=\rho_Q-2V=\rho_c\Omega_Q-2V$ and Eq.(\ref{potential}), we get
\begin{align}
p_Q=\rho_c\Omega_Q(-1+\Delta) .
\end{align}
Putting this relation into Eq. (\ref{a}), it becomes
\begin{align}
\frac{\ddot{a}}{a}=4\pi G\rho_c\left[\Omega_Q\left(1-\Delta\right)-\frac{1}{3}\Omega_Q\right]  .
\end{align}
As  $\dot{Q}^2=\rho_Q\Delta$, the first derivatives of $p_Q$ and $\rho_Q$ become
\begin{equation}
\dot{p}_Q=-3H\rho_Q\Delta-2V'\dot{Q}, \ \ \dot{\rho}_Q=-3H\rho_Q\Delta  .
\end{equation}
Using $H^2=(8\pi G/3)\rho_c$, the term on the right-hand side (\ \ ) of the second term in the [\ \ \ ] of Eq. (\ref{d^2w/da^2})  becomes 
\begin{align}
\ddot{a}\rho^2_Q+2\dot{a}\rho_Q\dot{\rho}_Q & = 4\pi Ga\rho_c\left[\Omega_Q\left(1-\Delta\right)-\frac{1}{3}\Omega_Q\right]\rho^2_Q+2Ha\rho_Q(-3H\rho_Q\Delta) \nonumber \\
 & = 4\pi Ga\rho^2_Q\rho_c\left[\Omega_Q\left(1-\Delta\right)-\frac{1}{3}\Omega_Q-4\Delta \right] . \label{4.42}
\end{align}
Taking $p_Q-\rho_Q=-2V$ and Eq.(\ref{4.33}) , the term on the left-hand side (\ \ ) of the second term  is 
\begin{align}
\dot{p}_Q\rho_Q-p_Q\dot{\rho}_Q = -2V\rho_Q\sqrt{\rho_c}\left[\sqrt{24\pi G}\Delta+\frac{V'}{V}\sqrt{\Omega_Q\Delta}\right]  .  \label{4.43}
\end{align}
From Eqs. (\ref{4.42}) and  (\ref{4.43}), the second term in the [\ \ \ ] of Eq. (\ref{d^2w/da^2}) becomes 
\begin{align}
(\dot{p}_Q\rho_Q-p_Q\dot{\rho}_Q) & (\ddot{a}\rho_Q^2+2\dot{a}\rho_Q\dot{\rho}_Q)
= -\frac{8\pi Ga\rho^5_Q\rho^{1/2}_c}{\Omega_Q}\left(1-\frac{\Delta}{2}\right) \nonumber \\
& \quad \times \left\{\left[\sqrt{24\pi G}\Delta+\frac{V'}{V}\sqrt{\Omega_Q\Delta}\right]\left[\Omega_Q\left(1-\Delta\right)-\frac{1}{3}\Omega_Q-4\Delta \right] \right\} .  \label{4.40}
\end{align}

Next, we try to calculate  the first term within [ \ \ \ ] of Eq. (\ref{d^2w/da^2}).  Equation (\ref{4.31}) can be changed to 
\begin{align}
\ddot{p}_Q\rho_Q-p\ddot{\rho}_Q  & = \left[-2\frac{V''}{V}\frac{\dot{Q}^2}{V}+6\frac{H\dot{Q}}{V}\frac{V'}{V}+2(\frac{V'}{V})^2 \right]\frac{1}{2}\dot{Q}^2V^2 \nonumber \\
& \quad +\left[(-6\frac{\ddot{a}}{a}\frac{1}{V}+42\frac{H^2}{V}-2\frac{V''}{V})\frac{\dot{Q}^2}{V}+18\frac{H\dot{Q}}{V}\frac{V'}{V}+2(\frac{V'}{V})^2\right]V^3 .\label{4.41}
\end{align}
Here we calculate the elements separately in the above equation as
\begin{align}
\frac{\dot{Q}^2}{V} & =\frac{\rho_Q\Delta}{\rho_Q(1-\Delta/2)}=\frac{\Delta}{1-\Delta/2}  ,\\
\frac{H\dot{Q}}{V} & =\frac{\sqrt{(8\pi G/3)\rho_c}\sqrt{\rho_c\Omega_Q\Delta}}{\rho_c\Omega_Q(1-\Delta/2)}=\frac{\sqrt{(8\pi G/3)\Delta}}{\sqrt{\Omega_Q}(1-\Delta/2)}  ,\\
\frac{1}{2}\dot{Q}^2V^2 & =\frac{1}{2}\rho_Q\Delta[\rho(1-\frac{\Delta}{2})]^2=\frac{1}{2}\rho^3_Q\Delta(1-\frac{\Delta}{2})^2  ,\\
\frac{\ddot{a}}{a}\frac{1}{V} & =\frac{4\pi G\rho_c\left[\Omega_Q\left(1-\Delta\right)-\frac{1}{3}\Omega_Q\right]}{\rho_c\Omega_Q(1-\Delta/2)} 
=\frac{4\pi G}{\Omega_Q}\frac{\left[\Omega_Q\left(1-\Delta\right)-\frac{1}{3}\Omega_Q\right]}{(1-\Delta/2)} , \\
\frac{H^2}{V} & =\frac{(8\pi G/3)\rho_c}{\rho_c\Omega_Q(1-\Delta/2)}=\frac{8\pi G}{3}\frac{1}{\Omega_Q(1-\Delta/2)}.
\end{align}
Then Eq. (\ref{4.41}) can be described as
\begin{align}
 & \ddot{p}_Q\rho_Q-p_Q\ddot{\rho}_Q= \frac{1}{2}\rho^3_Q\Delta^2\left(1-\frac{\Delta}{2}\right)\biggl[\left(-2-\frac{4}{\Delta}\left(1-\frac{\Delta}{2}\right)\right)\frac{V''}{V}  \nonumber \\
& \quad + \left(6\frac{\sqrt{8\pi G/3}}{\sqrt{\Omega_Q \Delta}}+\frac{36\sqrt{8\pi G/3}(1-\Delta/2)}{\Delta\sqrt{\Delta\Omega_Q}}\right)\frac{V'}{V}
+\left(\frac{2}{\Delta}\left(1-\frac{\Delta}{2}\right)+\frac{4}{\Delta^2}\left(1-\frac{\Delta}{2}\right)^2\right)\left(\frac{V'}{V}\right)^2 \nonumber \\
 & \qquad \qquad + \frac{2}{\Delta\Omega_Q}\left(-24\pi G\left(\Omega_Q\left(1-\Delta\right)-\frac{1}{3}\Omega_Q\right)+112\pi G\right) \biggr] .  \label{4.47}
\end{align}
As $\dot{a}\rho^2_Q=aH\rho^2_Q=a\sqrt{(8\pi/3)\rho_c}\rho^2_Q$, the first term within [ \ \ \ ] of Eq. (\ref{d^2w/da^2}) becomes
\begin{align}
 & (\ddot{p}_Q\rho_Q-p_Q\ddot{\rho}_Q)\dot{a}\rho^2_Q= \frac{1}{2}a\rho^5_Q\sqrt{(8\pi G/3)\rho_c}\Delta^2\left(1-\frac{\Delta}{2}\right)\biggl[\left(-\frac{4}{\Delta}\right)\frac{V''}{V} \nonumber \\
& + \left(6\frac{\sqrt{8\pi G/3}}{\sqrt{\Omega_Q \Delta}}+\frac{36\sqrt{8\pi G/3}(1-\Delta/2)}{\Delta\sqrt{\Delta\Omega_Q}}\right)\frac{V'}{V} 
 + \left(\frac{2}{\Delta}\left(1-\frac{\Delta}{2}\right)+\frac{4}{\Delta^2}\left(1-\frac{\Delta}{2}\right)^2\right)\left(\frac{V'}{V}\right)^2 \nonumber \\
  & \quad + \frac{2}{\Delta\Omega_Q}\left(-24\pi G\left(\Omega_Q\left(1-\Delta\right)-\frac{1}{3}\Omega_Q\right)+112\pi G\right) \biggr]  .   \label{4.48}
\end{align}

Then Eq. (\ref{d^2w/da^2}) can be written as 
\begin{align}
\frac{d^2w_Q}{da^2} & = \frac{1}{\dot{a}^3\rho^4_Q} \times \frac{1}{2}\sqrt{\frac{8\pi G}{3}}\rho^{1/2}_c\rho^5_Qa\Delta^2(1-\frac{\Delta}{2})
\biggl[([\, \  \ ]{\rm  in Eq. }(\ref{4.48}) ) \nonumber \\
& \quad +\frac{2\sqrt{24\pi G}}{\Omega_Q\Delta^2}(\{ \, \ \ \} {\rm in Eq. }(\ref{4.40}))\biggr] . \label{4.49}
\end{align}
The term within [\ \ \ \ ] is calculated as 
\begin{align}
& -\frac{4}{\Delta}\frac{V''}{V}+\frac{6}{\Delta^2}\sqrt{\frac{8\pi G}{3\Omega_Q} \Delta}\left((1-\Delta)(6+\Omega_Q)-\frac{1}{3}\Omega_Q\right)\frac{V'}{V} \nonumber \\
& \qquad +\left(\frac{2}{\Delta}\right)^2\left(1-\frac{\Delta}{2}\right)\left(\frac{V'}{V}\right)^2+\frac{16\pi G}{\Delta\Omega_Q}(14-12\Delta) .  \label{4.54}
\end{align}
The ouside factor of [\ \ \ \ ] is
\begin{align}
\frac{3}{16\pi G}\frac{\Omega_Q}{a^2}\left(1-\frac{\Delta}{2}\right)\Delta^2 .  \label{4.55}
\end{align}
Using Eqs. (\ref{4.54}) and (\ref{4.55}), $d^2w/da^2 $ becomes

\begin{align}
\frac{d^2w_Q}{da^2}= \frac{3\Omega_Q}{16\pi a^2}\left(1-\frac{\Delta}{2}\right)   &
 \biggl[-4\Delta M^2_{\rm pl}\frac{V''}{V}\label{4.56} +4\left(\frac{6\pi \Delta}{\Omega_Q}\right)^{0.5}\left(\left(1-\Delta\right)
 (6+\Omega_Q)-\frac{1}{3}\right)M_{\rm pl}\left(\frac{V'}{V}\right)\\
& \quad  +4(1-\frac{\Delta}{2})M^2_{\rm pl}\left(\frac{V'}{V}\right)^2+\frac{32\pi\Delta}{\Omega_Q}(7-6\Delta) \biggr] .  \nonumber 
\end{align}

\section{Matter-dominant and Attractor-solution Approximation}

For the era $\rho_M \geq \rho_Q$,  we can use Eq. (\ref{Qfield}) as the matter-dominant approximation where $a \propto t^{2/3}$.  For the inverse power potential, it becomes
\begin{align}
\ddot{Q}+\frac{2}{t}\dot{Q}-\alpha M^{4+\alpha} Q^{-{\alpha-1}}=0 ,  \label{Qfielda}
\end{align}
which has the attractor solution as \cite{16}  
\begin{align}
Q=\left(\frac{\alpha(2+\alpha)^2M^{4+\alpha}t^2}{2(4+\alpha)}\right)^{1/(2+\alpha)} . \label{Sol}
\end{align}
If this solution was used, the parameters $\alpha$ and $ M $ would be derived by $\Delta$ and the time $t_c$ at which $\rho_M=\rho_Q$.

For this solution, it is derived as
\begin{align}
\frac{1}{2}\dot{Q}^2/V=\frac{\alpha}{4+\alpha},
\end{align}
then
\begin{align}
w_Q=(\frac{\alpha}{4+\alpha}-1 )/(\frac{\alpha}{4+\alpha}+1)=-\frac{2}{2+\alpha}=-1+\frac{\alpha}{2+\alpha}.
\end{align}
As $\Delta=\frac{\alpha}{2+\alpha}$, $\alpha$ could be estimated by $\Delta$ as
\begin{align}
\alpha=\frac{2\Delta}{1-\Delta}.
\end{align}

If $a\simeq t^{2/3}$ and $\Delta \leq 0.1$ could be approximated until the current $t_0$, it could be approximated as $t_c=t_0/(\Omega_Q/(1-\Omega_Q))^{1/2}$.  
The parameter $M$ is derived by the relation $\rho_M=\rho_Q$ at $t_c$ as
\begin{align}
\frac{1}{2}\cdot \frac{1}{6\pi G t_c^2} & =M^{4+\alpha}\left(\frac{\alpha(2+\alpha)^2M^{4+\alpha}t_c^2}{2(4+\alpha)}\right)^{-\alpha/(2+\alpha)}  \label{Sola} \\
& =\left(\frac{2^{1+\alpha}(2+\alpha)^{2-\alpha}}{\alpha^{\alpha}(4+\alpha)^2}\right)^{1/(2+\alpha)}M^{2(4+\alpha)/(2+\alpha)}t_c^{-2\alpha/(2+\alpha)},
\end{align}
where $\rho_M=1/(6\pi G t^2)$ and $\rho_Q=\dot{Q}^2/2+V$ are used.  
So $M_{\ast}$ is determined by $\alpha$ and $t_c$ as
\begin{align}
M_{\ast}=\left((3 \cdot 2^3\pi)^{-(2+\alpha)}\alpha^{\alpha}(2+\alpha)^{-2+\alpha}(4+\alpha)^2\right)^{1/(2(4+\alpha))} \times t_c^{-2/(4+\alpha)}.
\end{align}
Then the observed values $t_0, \Omega_Q$, and $\Delta$ could determine the potential parameters $\alpha$ and $ M_{\ast} $ under the matter-dominant and attractor-solution approximation.
\vspace{1.5cm}

\hspace{-0.5cm}
\end{document}